# AI-generated art perceptions with GenFrame – an image-generating picture frame

Peter Kun[a*], Matthias Freiberger[b], Anders Sundnes Løvlie[a], Sebastian Risi[a]

[a] IT University of Copenhagen

[b] University of Copenhagen

*Corresponding author e-mail: peku@itu.dk

**Abstract**: Image-generation models are changing how we express ourselves in visual art. However, what people think of AI-generated art is still largely unexplored, especially compared to traditional art. In this paper, we present the design of an interactive research product, GenFrame – an image-generating picture frame that appears as a traditional painting but offers the viewer the agency to modify the depicted painting. In the current paper, we report on a study where we deployed the GenFrame in a traditional art museum and interviewed visitors about their views on AI art. When provoked by AI-generated art, people need more of the artist's backstory and emotional journey to make the artwork commensurate with traditional art. However, generative AI-enabled interactive experiences open new ways of engaging with art when a turn of a dial can modify art styles or motifs on a painting.
A demo can be seen here: https://youtu.be/1rhW4fazaBY.

**Keywords**: generative AI; image-generation models; museums and cultural heritage; research through design

## 1. Introduction

The emergence of AI image-generation models such as DALL-E, Midjourney, and Stable Diffusion (Rombach et al., 2022) has sparked substantial public interest as they started to reach professional-level quality in visual domains (Epstein et al., 2023; Roose, 2022b), such as illustration, stock photography, digital concept art (Cremer et al., 2023), interface design (Wei et al., 2023), or sketching (Lawton et al., 2023). Even a year ago, these powerful models were only accessible through complicated tooling; today, they are becoming ubiquitous, such as

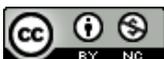

*Peter Kun, Matthias Freiberger, Anders Sundnes Løvlie, Sebastian Risi*

being embedded in search engines[1]. Along with the wide-spreading and powerful capabilities, a growing number of controversies are emerging around art and AI tools (Chen, 2023; Roose, 2022a); these models can imitate the styles of various artists from their extensive training sets and produce unlimited types of objects or domains. While these debates highlight misuse, we are still at the beginning to see what kind of positive and meaningful utilities these models could bring to our everyday lives.

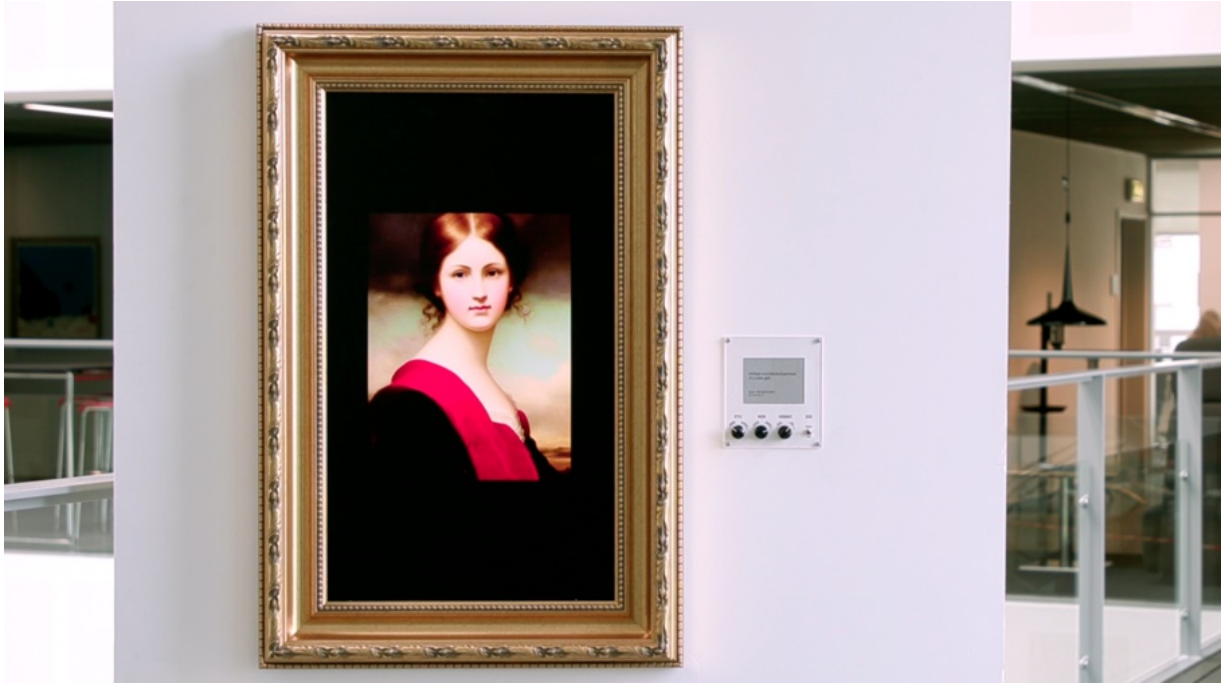

**Figure 1.** *The GenFrame.*

Situating generative image model capabilities within public contexts, such as traditional art museums, provides a promising direction. In the paper, we present the design of an image-generating picture frame – GenFrame, as shown in Figure 1 – which we deploy in a museum context. We draw inspiration from Krippendorff and Butter's (1984) work on product semantics; the GenFrame is designed to "speak" to the museum visitors, embodying the nuances expected in the context of a museum; the GenFrame is designed to imitate how traditional art is usually presented – in an ornate gold frame and a museum placard exhibit item next to it. However, the museum placard also serves as an interface where the viewer can modify the depicted painting.

We use the GenFrame to provoke museum visitors' views about AI-generated art to better understand future uses and potential interfaces for image-generation models in the context of museums and cultural heritage. The leading design question (DQ) and research questions (RQs) in the study are:

---

[1] Such as https://www.bing.com/chat, accessed on 28 October 2023.

- **DQ:** How can we embed generative image models into interactive artifacts?
- **RQ1:** How is AI-generated art perceived when presented in the context and embodiment of traditional art?
- **RQ2:** What do people think about image generation AI if they are made familiar with its possibilities in the context of museums and cultural heritage?

The paper makes two contributions. First, we provide a rare case study of embedding an image-generation model into an interactive artifact – the GenFrame. We report on the design considerations beyond regular "prompt engineering" commonly used today. Second, we use the GenFrame as a sensitizing tool for interviews with museum visitors. Through the interviews, we investigate perceptions around AI-generated art, especially in comparison to traditional art. The study's findings highlight that even if people like the results, they miss connections to the artist's backstory and lived experience expressed through an AI-generated artwork. Furthermore, people are conflicted about the authorship of AI-generated art, even when AI-generated art practices are potentially more communal and offer new ways of integrating AI into art practices. However, the GenFrame enabled an engaging experience, showing novel interaction design potential for image-generation models.

## 2. Related work

### 2.1 Image-generation models

The advent of the internet resulted in enormous collections of images annotated with accompanying related textual descriptions. Recently, significant progress in visual content generation has been achieved through advances in machine learning leveraging these collections (Radford et al., 2021; Ramesh et al., 2021; Vaswani et al., 2017) in combination with probabilistic generative approaches (Kingma & Welling, 2014; Sohl-Dickstein et al., 2015). Text-guided image-generation models, where the user specifies a textual description of the image they wish to see, like DALL-E, CLIP-VQGAN, and Stable Diffusion (Ramesh et al., 2021; Crowson et al., 2022; Rombach et al., 2022), have inspired a new wave of corporate, such as OpenAI and Midjourney, and open source, such as Stable Diffusion, research and development effort in image generation for creative tasks. An indicator of the progress made in a brief timespan is the many emerging techniques to convey more robust control over image generation: techniques like ControlNet (Zhang & Agrawala, 2023), LoRA (Hu et al., 2021), and Dreambooth (Ruiz et al., 2023) let users precisely determine how Stable Diffusion works. These help to better convey user intention - whether it be an image's content or style or things like people's poses or a content's contour sketch.

Such control is lower than what prompting, and prompt engineering provide. Prompting has become the primary interaction paradigm for generative models, offering an expressive interaction modality to harness the broad applicability of powerful models (Oppenlaender, 2022). However, prompting is problematic as an interface; unlike direct manipulation (Shneiderman, 1983), affording fine-grained control over tasks, prompting can yield unpredictable results, often resulting in brute-force techniques to convey a specific intent (Liu & Chilton, 2022). Alternative approaches have tried to address these issues by including initial images next to the text prompts (Qiao et al., 2022) or using rich-text (i.e., word formatting, color, style, or footnotes) as expressive prompt control (Ge et al., 2023). While these new techniques have been improving the overall prompting interaction paradigm, how to interact with these models when embedded into interactive experiences outside of the realms of end-user AI tools has yet to be explored.

### 2.2 AI-generated art

Image-generation models are not the first AI technologies used for making art. As outlined by several authors (Audry, 2021; Cetinic & She, 2022; Mazzone & Elgammal, 2019; Zylinska, 2020), people have experimented with art-making with computers for over 50 years, often by appropriating advancements in AI and machine learning, while continuously redefining what art means in the 21st century. Key turning point technologies were Generative Adversarial Networks (GANs, Goodfellow et al., 2014), where two neural networks compete to generate a realistic image. Along with more recent text-guided approaches, AI tools can increasingly reproduce the looks of traditional art, which has also generated critique (Chamberlain et al., 2018; McCormack et al., 2023). Nevertheless, with access to image-generation models and prompting's simplicity in interacting with these models, new artistic practices are developing, indicating the synthesis of AI technologies into traditional art (Caramiaux & Fdili Alaoui, 2022; Oppenlaender et al., 2023; Vartiainen & Tedre, 2023).

Paintings, in particular, have been one of the focuses of AI and automation art (Colton, 2012) and how the public perceives paintings and AI art (e.g., Lyu et al., 2022). With the recent popularity of image-generation models and their text-prompting practice, studies have discovered systematic ways to incorporate specific painting styles into prompting (Liu & Chilton, 2022) and whether artists use these tools differently than non-experts (Sun et al., 2022). Previous studies highlight a negative perception bias against AI (Chiarella et al., 2022), even when people hardly distinguish AI-generated paintings from human-made (Gangadharbatla, 2022; Park et al., 2023). However, people have a more favorable judgment of AI-generated art when a human is engaged in the making process (Bellaiche et al., 2023). Meanwhile, AI-generated art has generally been judged equivalent in artistic value (Hong & Curran, 2019). These studies highlight the controversies around people's perception and acceptance of AI-generated art, especially compared to human-made art.

## 3. Method

To address our research questions around AI-generated art perceptions in the context of museums and cultural heritage and our design agenda to investigate embedding generative image models into interactive artifacts, we employ a research-through-design (RTD) methodology (Stappers & Giaccardi, 2017; Zimmerman et al., 2007) for our inquiry's exploratory nature. Our design process is guided by Odom et al.'s (2016) conceptualization of research products; research products, going beyond research prototypes, feature high-quality finish to fit into their contexts to enable independent studies focused on inquiry-driven deployments. In our analysis, we are guided by Sengers and Gaver's (2006) view on accepting multiple interpretations of our design and the specific field study at a museum. Concretely, we use our design as a sensitizing tool (Sleeswijk Visser et al., 2005) to provoke people's broader perspectives on AI, art, and what these mean in the future.

### 3.1 GenFrame

GenFrame is an interactive picture frame consisting of a digital screen in a golden ornate frame and an interactive placard with a control panel (see Figure 1). The viewer interacts with GenFrame by modifying the dials and the switch on the interactive placard. Once interacted with, the digital screen refreshes to show a new image based on the settings on the placard. A demo can be viewed at https://youtu.be/1rhW4fazaBY. The following section elaborates on the design decisions that led to GenFrame's current embodiment.

Following Odom et al.'s (2016) view on a research product, we designed GenFrame to be inquiry-driven, contextually appropriate, and independent. These principles suggested making the aesthetic presentation resemble a museum exhibit item, using a traditional, golden ornate frame for the screen, and extending the placard seen next to a museum item for the interface. The viewer interacts with GenFrame by manipulating three dials and a switch, which control the style of the image, the mood of the depicted girl, and influence the AI model and seed, respectively. This controller interface was inspired by guitar pedals and musical instruments - allowing direct modification of specific parameters while retaining simplicity.

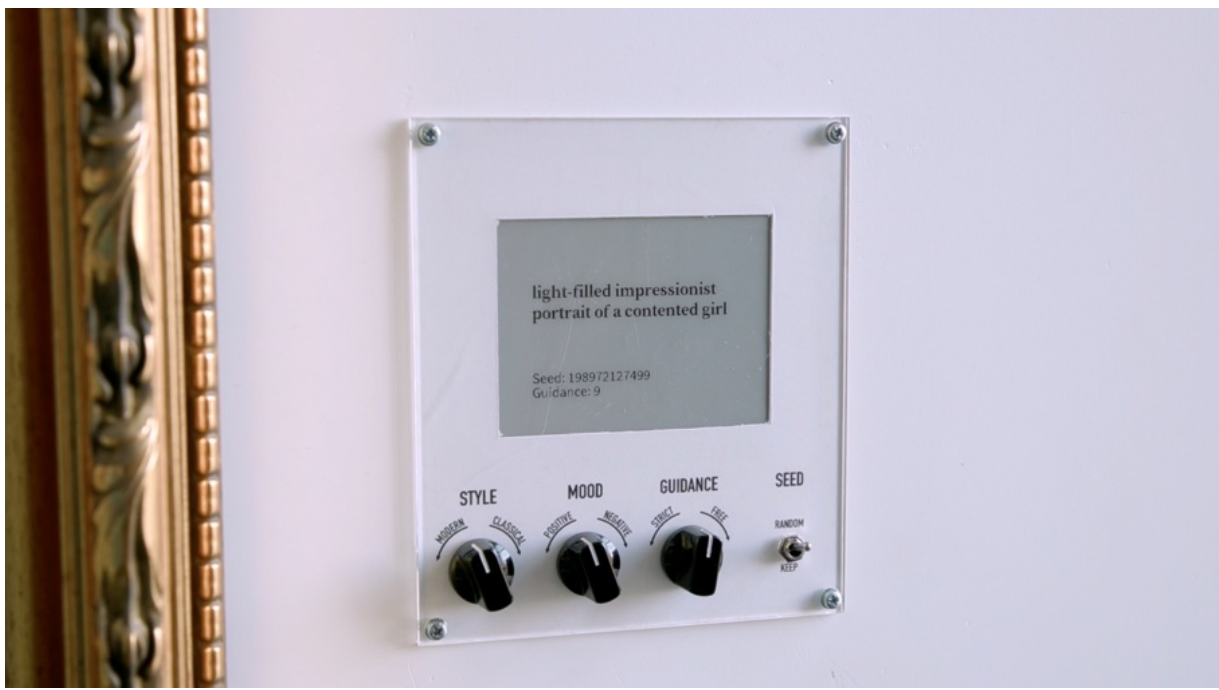

*Figure 2. The placard interface enables the viewer to change the painting.*

*Peter Kun, Matthias Freiberger, Anders Sundnes Løvlie, Sebastian Risi*

To make GenFrame independent and autonomous, we wanted to ensure complete control over the AI pipeline and alleviate black-boxing issues often associated with AI models (Zednik, 2021). Therefore, we deployed the standard Stable Diffusion V1.5 model without any fine-tuning on our server. While popular image generation services like Midjourney may offer a high-level aesthetic quality in general, Stable Diffusion, with customization techniques we employed as presented later, can achieve superior results. These tools enabled complete control over the process while remaining transparent and explicable. Due to computational limitations on our server, we needed to reduce the resolution of the image shown so as not to fill the whole screen (see Figure 5). While we rationalized this solution as showing the images in a "*passe-partout*", it is a distracting feature for some viewers.

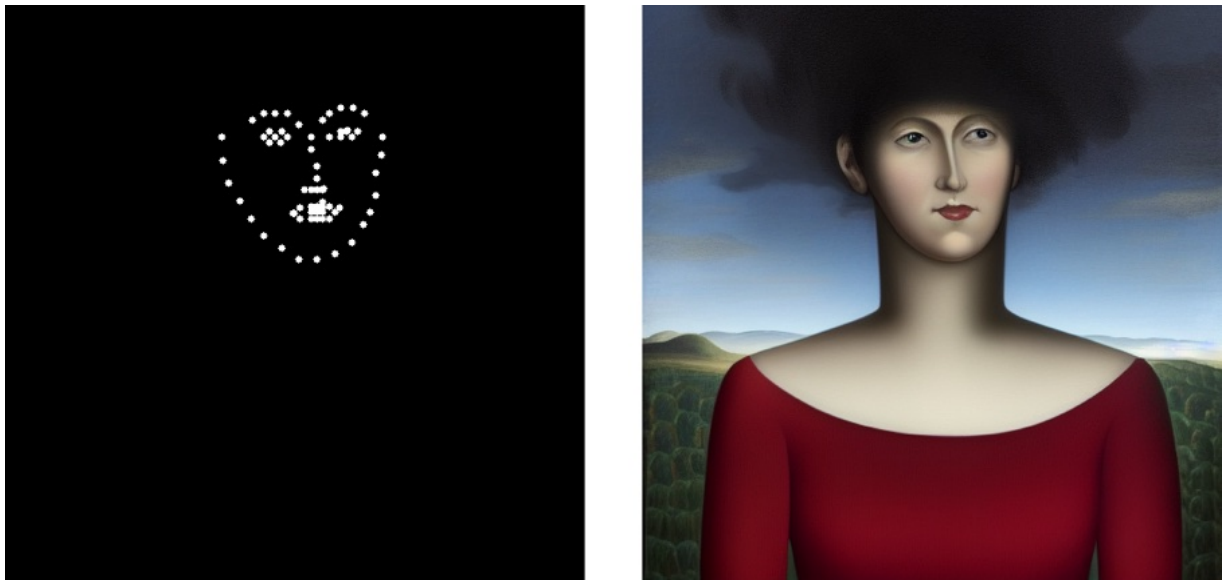

*Figure 3. The OpenPose face structure unifies all generated images.*

Different design iterations made a better viewer experience apparent when some elements were consistent between two generated paintings. To this end, we employed ControlNet (Zhang & Agrawala, 2023), an end-to-end neural network architecture allowing precise control over Stable Diffusion with task-specific conditions, using OpenPose (Cao et al., 2017) face-estimation. This feature ensured that despite the variability of the image generation, all images had a sense of cohesiveness: a generated face positioned at the exact locations (see Figure 3). With this design decision, we also unified all paintings to be portraits generated about "a girl" – a common trope throughout the history of canonical portrait paintings such as the Mona Lisa, the Girl with a Pearl Earring, etc.

We complemented the picture frame with a placard interface (Figure 2) to enable the viewer to manipulate the generated portrait of a girl. Through turning three dials, the viewer could manipulate the style of the painting, the mood of the girl, and the "classifier guidance" of the model; how closely the model followed its prompts would influence the quality of output (S. Hong et al., 2023). We also introduced a switch to set the seed the same or random;

keeping the same seed enables comparison between two generations of images to study how parameters modify the image.

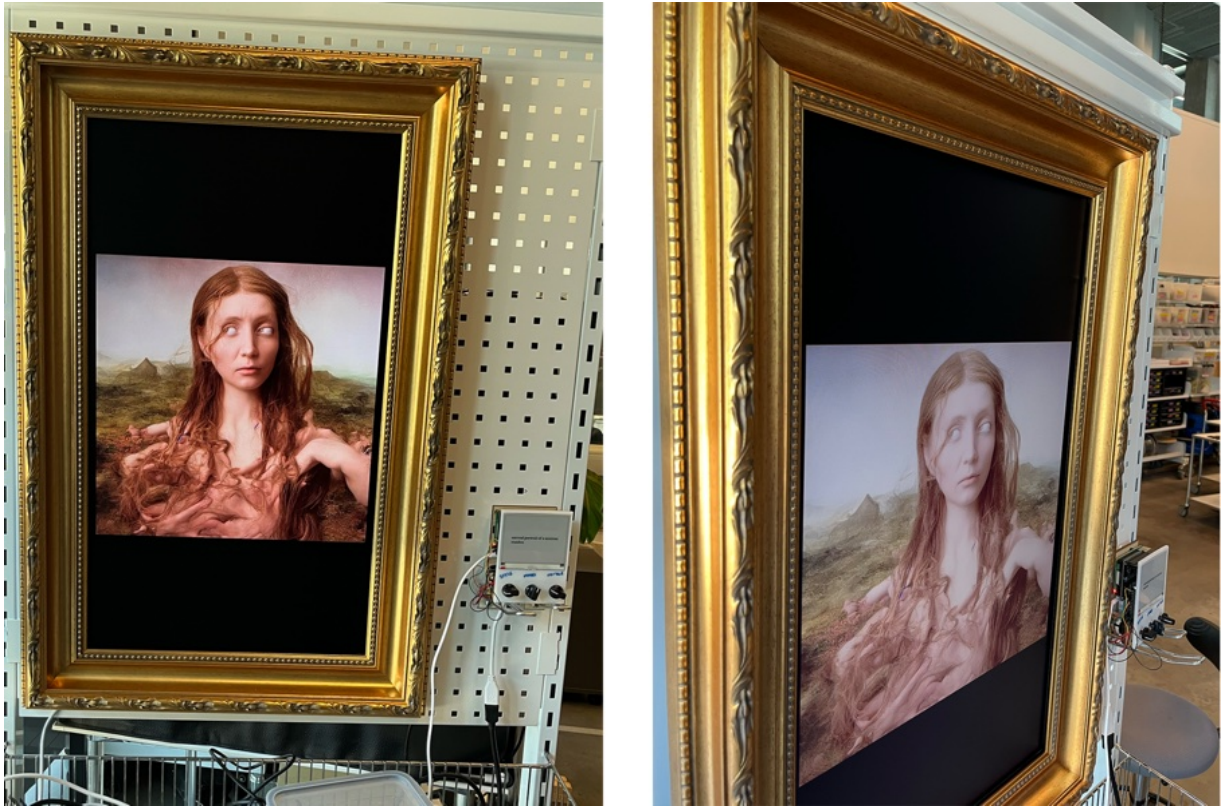

*Figure 4. Prototype view of the GenFrame. The screen's good side-viewing angles mimic the perception of a non-digital artwork even as viewers approach and look at it from different angles, as visitors do in a museum.*

We aimed for a high-quality finish commensurable to paintings and their context in museums. Figure 4 shows we achieved wide viewing angles and a physical painting-like appearance by using a matte Samsung The Frame 32" screen with an ornate golden frame emphasizing classical art aesthetics. In the placard interface, there is a Raspberry Pi as the core of the hardware, and connected to it are an e-ink display, three 12-state rotary switches, and a switch through an analog electronics board. The e-ink display was a crucial design decision to recreate the museum experience as much as possible through a realistic-looking placard. We developed custom software for the Raspberry Pi to monitor user input, handle logic and server communication, and showcase the images received on the screen. The system monitors the dials until the viewers stop interacting with them and then creates a new image. The e-ink display updates and the image generation take approximately fifteen seconds. In the meantime, the frame shows interim image versions masked by a Gaussian blur, so viewers see an image emerging smoothly out of the random noise.

*Peter Kun, Matthias Freiberger, Anders Sundnes Løvlie, Sebastian Risi*

*3.2 Field study*

**Setup:** We deployed the GenFrame for a field study in the Danish National Gallery (SMK) foyer for one day at the end of August 2023. Passer-by museum visitors could interact and engage with the GenFrame (Figure 5). One researcher was present who invited people for an interview after interacting with the GenFrame.

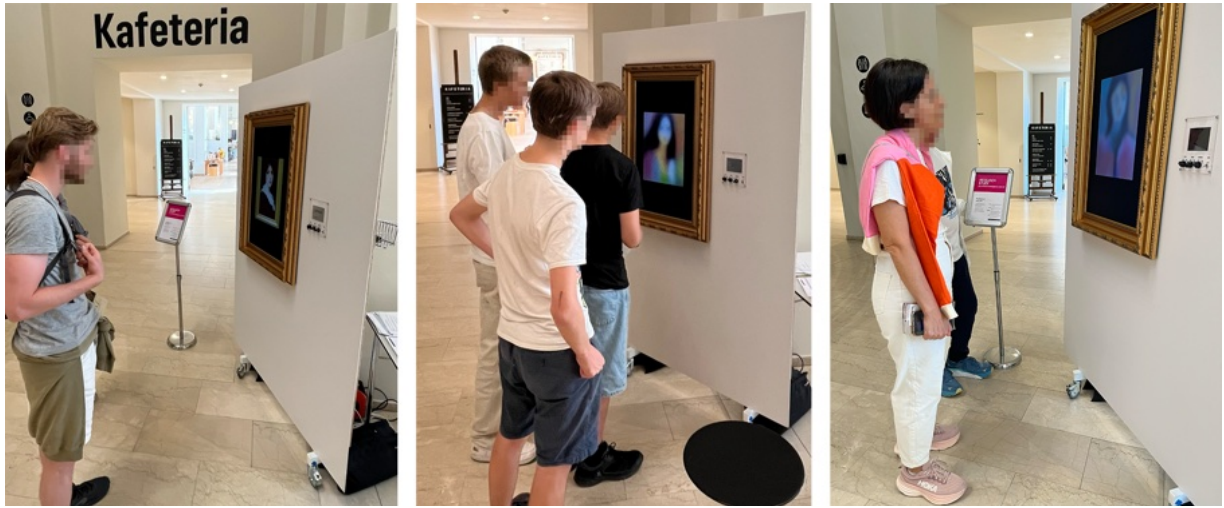

*Figure 5. Museum visitors are interacting with the GenFrame.*

**Participants:** Eighteen people (7 female, 11 male) participated in our interviews, with a mean age of 34.4 (SD = 19.4) and a mean number of museum visits in the past 12 months 6.1 (SD = 5.1). The participants came from 8 countries, as shown in Table 1. The researcher approached participants after interacting with the GenFrame without incentives (besides refreshments offered). All participants were museum visitors who were unknown to us beforehand.

*Table 1   Study participants.*

| Participant code | Age | Gender | Nationality | Museum visits in last 12 months | Profession |
|---|---|---|---|---|---|
| P01 | 24 | Female | French | 10 | PhD student in psychology |
| P02 | 24 | Male | French | 10 | business and entrepreneurship |
| P03 | 70 | Female | French | 5 | retired |
| P04 | 38 | Female | German | 3 | secondary school teacher |
| P05 | 76 | Male | Danish | 10 | retired |
| P06 | 26 | Female | Turkey | 20 | creative technologist, AI artist |

| P07 | 41 | Female | British | 10 | photographer |
| --- | --- | --- | --- | --- | --- |
| P08 | 41 | Male | Argentinian | 4 | N/A |
| P09 | 14 | Male | Denmark | 3 | high school student |
| P10 | 15 | Male | Denmark | 2 | high school student |
| P11 | 14 | Male | Turkey | 4 | high school student |
| P12 | 57 | Female | German | 1 | N/A |
| P13 | 19 | Male | Australian | 11 | student |
| P14 | 16 | Male | French | 1 | high school student |
| P15 | 35 | Male | Swedish | 4 | marketing consultant |
| P16 | 58 | Male | British | 10 | cybersecurity consultant |
| P17 | 25 | Male | Danish | 1 | higher education student |
| P18 | 26 | Female | Danish | 1 | higher education student |

**Data collection and analysis:** We conducted semi-structured interviews with the interested museum viewers who have interacted with the GenFrame. At the beginning of each interview, participants were informed about the data collected and signed an information statement in accordance with the university's policies and the General Data Protection Regulation (GDPR). Participants were asked to complete a form with basic demographic questions on age, gender (self-declared), nationality, and number of museum visits in the last 12 months. Table 1 summarizes these, with additional data we extracted from the interviews on current professions. The interviews' average duration was 11 minutes and 42 seconds. The interview guide was informed by 1) our research questions to evaluate the GenFrame, 2) eliciting reflections on the potential role of AI-generated art in the context of museums, and 3) the asking for perspectives on AI in general. We also collected the participants' current experience with AI and, if relevant, their current profession, which could influence their perspectives on AI. The interviews were audio recorded and transcribed verbatim by the leading researcher. We analyzed the transcribed interviews through thematic analysis, following the guidelines by Braun and Clarke (2006).

## 4. Results

*4.1 Assessing the GenFrame Experience*
**Playfulness:** The GenFrame was frequently described as a playful and engaging experience, particularly when it allowed social interaction. "Fun, creative, especially when you are with somebody else," one participant described, emphasizing the shared engagement it provided

*Peter Kun, Matthias Freiberger, Anders Sundnes Løvlie, Sebastian Risi*

(P15). This sense of play extended to the ability to manipulate the emotional tone of the artwork. Participants found amusement in altering the portrait's mood, with one recounting, "I found it amusing to change the mood of the portrait, making it look either upset or happy" (P12). The mood dial added an unexpected layer of interactivity, offering a novel form of engagement with the artwork that traditional pieces do not afford.

**Make-Believe and Analog Interaction:** Participants were drawn to the make-believe aspect of the GenFrame, noting how the experience allowed them to step into an artist's shoes. People found blending traditional and modern, reality and digital artifice fascinating. One participant likened the experience to mixing primary colors, saying, "What I liked most was the ability to simply turn dials and see where that combination takes you. It is a bit like mixing colors" (P13). This tangible, analog interaction was a notable contrast to the touchscreens dominating current technological interactions. "I love how analog it is; it feels quite old school," another participant shared (P07). As one participant eloquently said, the dials became instruments of creativity: "The dials act as my pen strokes, albeit much easier to manage than manually making a thousand pen strokes" (P15).

## 4.2 Does it make you an artist?

**Conflicting views on authorship:** Experiencing the GenFrame sparked conflicting views among participants about what it means to be an artist and how technology fits into this picture. Some felt the frame challenged the idea of who the "real" artist was, as one participant said, "Perhaps I am the artist. It [the GenFrame] raises questions about authorship when you can modify existing work so easily" (P01). However, the positive sentiment was not universal. Many felt held back by the GenFrame's limits, with one sharing, "I do not feel like the artist because of the limited parameters" (P06). They wanted to dive deeper, beyond just turning knobs.

**Limited control:** While participants found the GenFrame intriguing, many expressed a sense of detachment, mainly due to their limited influence over the final artistic output. The limits of control seemed to diminish their sense of contribution, with one participant stating, "I feel I do not have an actual impact on how it is going to look like. If I cannot influence the outcome, then I cannot claim to be the artist" (P12). This sentiment was mirrored by others who felt the simplicity of interaction reduced their role to that of bystanders: "It is like anyone could twist this, this, and this, and then something comes out. It is almost like just hitting the random button and seeing what happens, in a way" (P18).

**Too easy:** Another participant highlighted the triviality of the process, implying a lack of uniqueness or personal touch, saying, "It feels like anyone could do it—you know, a child, even a monkey, for example. So, I do not really feel like I played a part in it" (P17). This feedback suggests a more hands-on artistic process where users can witness and feel responsible for the transformation, thereby feeling more connected to the result. The current form of GenFrame seemed to offer more of a spectacle than an interactive art-making experience, leaving participants feeling unfulfilled in their creative expression.

## 4.3 "Is this art?"

**Emotions as a vehicle for art:** To provoke the participants' views on how they would frame something as art, we asked an intentionally ambiguous question to all of them while pointing at the GenFrame: Is this art? The participants could interpret "this" as the GenFrame itself or the AI-generated portraits on it, and often, they reflected deeper on "what is art, really?". The participants' reactions revealed deep-seated beliefs about the essence of artistic expression. These beliefs centered around conveying an artist's lived experiences and emotions, the artist's historical context, and the intrinsic value of creating something unprecedented and reflective of skilled craftsmanship. In its truest form, many participants felt art is deeply human, expressing the artist's emotions and messages. Participants commented that artists' personal experiences make the essence of art an "emotional and intellectual investment" (P05), which dilutes with AI-generated art. One stated, "There are no human emotions in the painting. The artist can paint her emotions and send a message" (P01), highlighting a perceived lack of emotional depth in AI-generated art. Others echoed this sentiment and found AI lacking the "spark of creativity that really makes you feel something" (P07), suggesting art's true resonance lies in its ability to embody and convey human emotion.

**Broader definition of art:** Conversely, some participants argued for a broader definition of art, including new forms of creation and interaction. "The relationship between what has been created and how the audience interacts with it contributes to the idea of what art can be" (P13), one participant contended, suggesting artistry might also reside in the experience generated between the viewer and the piece. Others questioned the notion of originality in art, pondering whether true creativity arises from within or merely reflects external influences: "It is doing something creative in its own way. But then again, because it is all based on pre-existing references [the training data], it also feels like, hmm, is it truly original?" (P18). Despite these differing views, there was an admiration for traditional artists' historical context and personal backstory. Participants expressed classical art serving as a "slice of history" (P17) and resonating differently because it "carries a lot of history with it" (P17); this historical weight appears to add significance and value, enriching the viewer's experience. In contrast, the GenFrame often lacked this rich historical context and personal touch, reducing the experience to something more akin to viewing a digital screen at home. Ultimately, these reflections underscore the importance of the artist's journey and the historical context in crafting pieces that resonate deeper beyond aesthetics.

## 4.4 AI-art as a Communal Future Craft

**Interactive art:** Participants thought the GenFrame counts more as tech or interactive art than comparable to traditional paintings. "I think the interactive aspect adds a crucial dimension. The relationship between what has been created and how the audience interacts with it contributes to the idea of what art can be" (P13), one participant argued, signifying the audience interaction in shaping the art piece. However, there was also an awareness of the trade-offs involved, as another participant remarked, "It is just another form of technology art. It might lack emotional depth, but it is still art in a different sense" (P02). This reflects that while interactive tech art diverges from traditional emotional expressiveness, it compensates by offering novel forms of engagement and representation.

By focusing on these first-hand accounts, we observe a cautious yet optimistic recognition among participants of AI and interactive technology's place in the artistic landscape. These emerging art forms are seen not as replacements for traditional art but as novel expansions of the artistic realm, offering new avenues for collective storytelling and audience participation.

**Communal experience:** Participants indicated AI art fosters a sense of collective engagement, differing significantly from traditional art's individualistic creation process. One participant – who works in creative coding and does AI art – noted the collective aspect of AI art, stating, "When I create AI art, it feels more collective; I have access to a lot of data from other humans at once" (P06). This perspective highlights an essential attribute of AI art: its ability to encapsulate and reflect a broader range of human experiences drawn from extensive data rather than a single artist's viewpoint. The notion of shared ownership was also evident, with reflections on the interactive nature of the artwork and the fact that any viewer can change it. As one participant described, "You create something yourself, but then when others interact with it, they sort of take on a bit of ownership as well" (P13), indicating a shift from art as a static piece to a dynamic, communal experience.

## 5. Discussion

### 5.1 Embedding Image-generation Models into Interactive Artifacts

To address our design question (DQ) on embedding generative image models into interactive artifacts, reflecting on the design and development process of the GenFrame revealed both opportunities and challenges in integrating AI. In our design process, it became clear that *controlling* an image-generation model is crucial to embedding them in an interactive artifact. Under control, we refer to "delimiting" an image-generation model's broad capabilities towards a specific design intent and ensuring all image generations are meaningful for the context. While ControlNet (Zhang & Agrawala, 2023) and OpenPose (Cao et al., 2017) served us to delimit from the AI pipeline, the design decisions around what the interface dials influence were just as significant. One way to look at the problem of delimiting image-generation models' broad capabilities is to approach it as an exploration of a parameter space (van Wijk & van Overveld, 2003). In this way, we considered each dial to reduce the dimensions of options and modify something 1) intentional in the context – the painting style; 2) surprising, provoking – changing the girl's mood; 3) related to the technology – to influence the image-generation model. While AI guidance was the least interesting, viewers found mood a playful way to influence the painting, and changing the styles was captivating and relevant. However, interviewees found the dials' dimension reductions limiting to feel like an artist. They claimed that additional dimensions – additional dials or a keyboard with open-ended text input – could improve the feel of ownership.

To conclude, the design challenge in embedding image-generation models is balancing sufficient control for the user and maintaining the model's capacity for creativity and novelty.

### *5.1 Perceptions of AI-generated Art in the Museum*

Regarding our first research question (RQ1) on museum visitors' perceptions of AI-generated art when embodied as traditional art, people viewed AI-generated art as exciting and appealing as a novel technology. However, people also held contradictory views about lacking emotion and not necessarily viewing it as art commensurate to traditional art. A critical perception that emerged was the importance of conveying lived experiences and emotions for art. Many interview participants tied the essence of art to the artists' journey and investment. This sentiment aligns with the philosophy of Merleau-Ponty (1964), who emphasized the embodied experience of creating and perceiving art as a connection between perception, body, and the world. Creating AI-generated paintings with the GenFrame seemed to dilute the artistic process' emotional depth and personal resonance for some participants. The absence of a backstory and context for the artwork's creation diminished the sense of a unique personal touch. While this notion stands for the current study, in other AI-art (Zylinska, 2020), there is artistic craftsmanship (Caramiaux & Fdili Alaoui, 2022) and artistic composition (Lyu et al., 2022) present.

Nevertheless, while the participants viewed the GenFrame's AI-generated art potentially through a negative bias (Chiarella et al., 2022), people were excited about the playful experience and the educational aspects of learning about art by quickly exploring various painting styles through the turn of a dial. A school teacher participant was enthusiastic about teaching art history through an artifact like the GenFrame, showing educational utility for the technology without requiring it to be viewed as an art-generating device.

To conclude, people expect to perceive the artist's labor and emotions from art, even if it is AI-generated. From a design perspective, reducing too much friction from generating art makes it less likely to be perceived as art.

### *5.2 AI and Art Futures*

Regarding the second research question on museum visitors' thoughts around generative AI, people reported a particular communal creativity process that comes into play through AI, which is remarkably different from traditional, individualistic artistic processes. Such a communal creation aspect with AI comes from multiple directions: P06 – with a personal AI prompt artistic practice – reflected on the communal aspects of AI art because of how the whole communities of practice online have emerged, where people learn prompting techniques from each other. Others mentioned a communal connection to all the artworks in the training data and that the image-generation model was synthesizing them. In a practical aspect, P15 reflected on the communal, social interactions with the GenFrame, where viewers can change the depicted artwork together rather than passively viewing it. The communal interaction enabled a shared experience of modifying the art rather than passive viewing.

While the co-creational aspects of practice are well-documented in recent literature (Caramiaux & Fdili Alaoui, 2022; Oppenlaender et al., 2023), communal aspects through the training data of the image-generation model means a humanizing perspective to connect to the different artists making up the dataset.

Participants mostly agreed that art practice would synthesize AI as a technique, just as it has incorporated previous technological progress. This was emphasized by the need for the artist's backstory and conveyed emotional state – to make that opaque, curation remains a crucial function of museums in guiding the public's appreciation of AI-infused artistic practices.

To conclude, art is expected to have AI-generated aspects in the future, and human curation remains essential. Compared to previous technology leaps, the communal aspects of AI-generated art might be more emphasized because the world has become globally connected through the internet.

*5.5 Limitations and Future Work*

This study was an initial exploration with limitations on small sample size and single site deployment that may reduce generalizability. The provoking aspects of the GenFrame in a museum led to perspectives influenced by novelty bias and could be altered through longitudinal and multi-site studies. We also acknowledge that showing the "portrait of a girl" is an overly gendered perspective; while we rationalize it to blend into the museum context, there are more progressive views on gender in contemporary culture. We plan a future experiment with a dial that changes the portrait between binary genders and explores the limits of image-generation models and how they can be controlled to depict non-binary people. Furthermore, as the public's opinions around AI-generated art remain relevant to study, future studies hint towards additional studies beyond a museum. Future work can also build on the qualitatively explored themes in the current work and aim to identify generalizable learnings through quantitative methods.

# 6. Conclusions

This paper presented the GenFrame, an interactive image-generating picture frame, and findings from its deployment provoking public perceptions of AI-generated art. The study revealed conflicting views on AI-generated art in a museum context. Participants were excited by GenFrame's novel experience but felt it lacked traditional art's emotional resonance and backstory. This highlights the importance of conveying the artist's journey and lived experiences, even in AI-assisted art. While AI-generated art was seen as interactive and communal, traditional art better conveys individual expression. The work shows design challenges in balancing meaningful agency for the user while controlling the broad capabilities of an image-generation model. Findings can inform the human-AI interaction design practice and inform practitioners working on AI-generated art to enhance authorial control and meaning.

**Acknowledgments:** This work was supported by a research grant (40575) from VILLUM FONDEN. We thank Jonas Heide Smith and Vivian Anholm at the National Gallery of Denmark for their help and support for running the field study. We also would like to acknowledge the work of Halfdan Hauch Jensen and the ITU AIR Lab in their support of developing our research prototype.

## 7. References

Audry, S. (2021). *Art in the age of machine learning*. MIT Press.

Bellaiche, L., Shahi, R., Turpin, M. H., Ragnhildstveit, A., Sprockett, S., Barr, N., Christensen, A., & Seli, P. (2023). Humans versus AI: Whether and why we prefer human-created compared to AI-created artwork. *Cognitive Research: Principles and Implications*, *8*(1), 42. https://doi.org/10.1186/s41235-023-00499-6

Braun, V., & Clarke, V. (2006). Using thematic analysis in psychology. *Qualitative Research in Psychology*, *3*(2), 77–101. https://doi.org/10.1191/1478088706qp063oa

Cao, Z., Simon, T., Wei, S.-E., & Sheikh, Y. (2017). *Realtime Multi-Person 2D Pose Estimation Using Part Affinity Fields*. 7291–7299. https://openaccess.thecvf.com/content_cvpr_2017/html/Cao_Realtime_Multi-Person_2D_CVPR_2017_paper.html

Caramiaux, B., & Fdili Alaoui, S. (2022). "Explorers of Unknown Planets": Practices and Politics of Artificial Intelligence in Visual Arts. *Proceedings of the ACM on Human-Computer Interaction*, *6*(CSCW2), 477:1-477:24. https://doi.org/10.1145/3555578

Cetinic, E., & She, J. (2022). Understanding and Creating Art with AI: Review and Outlook. *ACM Transactions on Multimedia Computing, Communications, and Applications*, *18*(2), 66:1-66:22. https://doi.org/10.1145/3475799

Chamberlain, R., Mullin, C., Scheerlinck, B., & Wagemans, J. (2018). Putting the art in artificial: Aesthetic responses to computer-generated art. *Psychology of Aesthetics, Creativity, and the Arts*, *12*(2), 177–192. https://doi.org/10.1037/aca0000136

Chen, M. (2023). Artists and Illustrators Are Suing Three A.I. Art Generators for Scraping and "Collaging" Their Work Without Consent. *Artnet News*. https://news.artnet.com/art-world/class-action-lawsuit-ai-generators-deviantart-midjourney-stable-diffusion-2246770

Chiarella, S. G., Torromino, G., Gagliardi, D. M., Rossi, D., Babiloni, F., & Cartocci, G. (2022). Investigating the negative bias towards artificial intelligence: Effects of prior assignment of AI-authorship on the aesthetic appreciation of abstract paintings. *Computers in Human Behavior*, *137*, 107406. https://doi.org/10.1016/j.chb.2022.107406

Colton, S. (2012). The Painting Fool: Stories from Building an Automated Painter. In J. McCormack & M. d'Inverno (Eds.), *Computers and Creativity* (pp. 3–38). Springer Berlin Heidelberg. https://doi.org/10.1007/978-3-642-31727-9_1

Cremer, D. D., Bianzino, N. M., & Falk, B. (2023, April 13). How Generative AI Could Disrupt Creative Work. *Harvard Business Review*. https://hbr.org/2023/04/how-generative-ai-could-disrupt-creative-work

Epstein, Z., Hertzmann, A., Herman, L., Mahari, R., Frank, M. R., Groh, M., Schroeder, H., Smith, A., Akten, M., Fjeld, J., Farid, H., Leach, N., Pentland, A., & Russakovsky, O. (2023). *Art and the science of generative AI: A deeper dive* (arXiv:2306.04141). arXiv. https://doi.org/10.48550/arXiv.2306.04141

Gangadharbatla, H. (2022). The Role of AI Attribution Knowledge in the Evaluation of Artwork. *Empirical Studies of the Arts*, *40*(2), 125–142. https://doi.org/10.1177/0276237421994697

Goodfellow, I., Pouget-Abadie, J., Mirza, M., Xu, B., Warde-Farley, D., Ozair, S., Courville, A., & Bengio, Y. (2014). Generative Adversarial Nets. In Z. Ghahramani, M. Welling, C. Cortes, N. Lawrence,

& K. Q. Weinberger (Eds.), *Advances in Neural Information Processing Systems* (Vol. 27). Curran Associates, Inc. https://proceedings.neurips.cc/paper_files/paper/2014/file/5ca3e9b122f61f8f06494c97b1afccf3-Paper.pdf

Hong, J.-W., & Curran, N. M. (2019). Artificial Intelligence, Artists, and Art: Attitudes Toward Artwork Produced by Humans vs. Artificial Intelligence. *ACM Transactions on Multimedia Computing, Communications, and Applications*, *15*(2s), 58:1-58:16. https://doi.org/10.1145/3326337

Hong, S., Lee, G., Jang, W., & Kim, S. (2023). Improving Sample Quality of Diffusion Models Using Self-Attention Guidance. *Proceedings of the IEEE/CVF International Conference on Computer Vision (ICCV)*, 7462–7471.

Kingma, D. P., & Welling, M. (2014). Auto-Encoding Variational Bayes. *Stat*, *1050*, 1.

Krippendorff, K., & Butter, R. (1984). Product Semantics: Exploring the Symbolic Qualities of Form. *Innovation-the European Journal of Social Science Research*, *3*, 4.

Lawton, T., Ibarrola, F. J., Ventura, D., & Grace, K. (2023). Drawing with Reframer: Emergence and Control in Co-Creative AI. *Proceedings of the 28th International Conference on Intelligent User Interfaces*, 264–277. https://doi.org/10.1145/3581641.3584095

Liu, V., & Chilton, L. B. (2022). Design Guidelines for Prompt Engineering Text-to-Image Generative Models. *Proceedings of the 2022 CHI Conference on Human Factors in Computing Systems*, 1–23. https://doi.org/10.1145/3491102.3501825

Lyu, Y., Wang, X., Lin, R., & Wu, J. (2022). Communication in Human–AI Co-Creation: Perceptual Analysis of Paintings Generated by Text-to-Image System. *Applied Sciences*, *12*(22), Article 22. https://doi.org/10.3390/app122211312

Mazzone, M., & Elgammal, A. (2019). Art, Creativity, and the Potential of Artificial Intelligence. *Arts*, *8*(1), Article 1. https://doi.org/10.3390/arts8010026

McCormack, J., Cruz Gambardella, C., Rajcic, N., Krol, S. J., Llano, M. T., & Yang, M. (2023). Is Writing Prompts Really Making Art? In C. Johnson, N. Rodríguez-Fernández, & S. M. Rebelo (Eds.), *Artificial Intelligence in Music, Sound, Art and Design* (pp. 196–211). Springer Nature Switzerland. https://doi.org/10.1007/978-3-031-29956-8_13

Merleau-Ponty, M. (1964). Eye and Mind. In *The primacy of perception: And other essays on phenomenological psychology, the philosophy of art, history, and politics*. Northwestern University Press.

Odom, W., Wakkary, R., Lim, Y., Desjardins, A., Hengeveld, B., & Banks, R. (2016). From Research Prototype to Research Product. *Proceedings of the 2016 CHI Conference on Human Factors in Computing Systems*, 2549–2561. https://doi.org/10.1145/2858036.2858447

Oppenlaender, J., Linder, R., & Silvennoinen, J. (2023). *Prompting AI Art: An Investigation into the Creative Skill of Prompt Engineering* (arXiv:2303.13534). arXiv. https://doi.org/10.48550/arXiv.2303.13534

Park, J., Kang, H., & Kim, H. Y. (2023). Human, Do You Think This Painting is the Work of a Real Artist? *International Journal of Human–Computer Interaction*, *0*(0), 1–18. https://doi.org/10.1080/10447318.2023.2232978

Radford, A., Kim, J. W., Hallacy, C., Ramesh, A., Goh, G., Agarwal, S., Sastry, G., Askell, A., Mishkin, P., Clark, J., Krueger, G., & Sutskever, I. (2021). Learning Transferable Visual Models From Natural Language Supervision. *Proceedings of the 38th International Conference on Machine Learning*, 8748–8763. https://proceedings.mlr.press/v139/radford21a.html

Ramesh, A., Pavlov, M., Goh, G., Gray, S., Voss, C., Radford, A., Chen, M., & Sutskever, I. (2021). Zero-shot text-to-image generation. *International Conference on Machine Learning*, 8821–8831.

Rombach, R., Blattmann, A., Lorenz, D., Esser, P., & Ommer, B. (2022). *High-Resolution Image Synthesis with Latent Diffusion Models* (arXiv:2112.10752). arXiv. https://doi.org/10.48550/arXiv.2112.10752

Roose, K. (2022a, September 2). An A.I.-Generated Picture Won an Art Prize. Artists Aren't Happy. *The New York Times*. https://www.nytimes.com/2022/09/02/technology/ai-artificial-intelligence-artists.html

Roose, K. (2022b, October 21). A.I.-Generated Art Is Already Transforming Creative Work. *The New York Times*. https://www.nytimes.com/2022/10/21/technology/ai-generated-art-jobs-dall-e-2.html

Sengers, P., & Gaver, B. (2006). Staying open to interpretation: Engaging multiple meanings in design and evaluation. *Proceedings of the 6th Conference on Designing Interactive Systems*, 99–108. https://doi.org/10.1145/1142405.1142422

Sleeswijk Visser, F., Stappers, P. J., Lugt, R. van der, & Sanders, E. B.-N. (2005). Contextmapping: Experiences from practice. *CoDesign*, *1*(2), 119–149. https://doi.org/10.1080/15710880500135987

Sohl-Dickstein, J., Weiss, E., Maheswaranathan, N., & Ganguli, S. (2015). Deep Unsupervised Learning using Nonequilibrium Thermodynamics. *Proceedings of the 32nd International Conference on Machine Learning*, 2256–2265. https://proceedings.mlr.press/v37/sohl-dickstein15.html

Stappers, P. J., & Giaccardi, E. (2017). Research through Design. In M. Soegaard & R. Friis-Dam (Eds.), *The Encyclopedia of Human-Computer Interaction* (2nd ed., pp. 1–94). The Interaction Design Foundation.

Sun, Y., Yang, C.-H., Lyu, Y., & Lin, R. (2022). From Pigments to Pixels: A Comparison of Human and AI Painting. *Applied Sciences*, *12*(8), Article 8. https://doi.org/10.3390/app12083724

van Wijk, J. J., & van Overveld, C. W. A. M. (2003). Preset Based Interaction with High Dimensional Parameter Spaces. In F. H. Post, G. M. Nielson, & G.-P. Bonneau (Eds.), *Data Visualization: The State of the Art* (pp. 391–406). Springer US. https://doi.org/10.1007/978-1-4615-1177-9_27

Vartiainen, H., & Tedre, M. (2023). Using artificial intelligence in craft education: Crafting with text-to-image generative models. *Digital Creativity*, *34*(1), 1–21. https://doi.org/10.1080/14626268.2023.2174557

Vaswani, A., Shazeer, N., Parmar, N., Uszkoreit, J., Jones, L., Gomez, A. N., Kaiser, Ł., & Polosukhin, I. (2017). Attention is All you Need. *Advances in Neural Information Processing Systems*, *30*. https://proceedings.neurips.cc/paper_files/paper/2017/hash/3f5ee243547dee91fbd053c1c4a845aa-Abstract.html

Wei, J., Courbis, A.-L., Lambolais, T., Xu, B., Bernard, P. L., & Dray, G. (2023). *Boosting GUI Prototyping with Diffusion Models* (arXiv:2306.06233). arXiv. https://doi.org/10.48550/arXiv.2306.06233

Zednik, C. (2021). Solving the Black Box Problem: A Normative Framework for Explainable Artificial Intelligence. *Philosophy & Technology*, *34*(2), 265–288. https://doi.org/10.1007/s13347-019-00382-7

Zhang, L., & Agrawala, M. (2023). *Adding Conditional Control to Text-to-Image Diffusion Models* (arXiv:2302.05543). arXiv. https://doi.org/10.48550/arXiv.2302.05543

Zimmerman, J., Forlizzi, J., & Evenson, S. (2007). Research through design as a method for interaction design research in HCI. *Proceedings of the SIGCHI Conference on Human Factors in Computing Systems*, 493–502. https://doi.org/10.1145/1240624.1240704

Zylinska, J. (2020). *AI Art: Machine Visions and Warped Dreams*. Open Humanities Press. http://www.openhumanitiespress.org/books/titles/ai-art/

About the Authors:

**Peter Kun** is a Postdoctoral Researcher at the Media, Arts and Design group at the IT University of Copenhagen. His research agenda on meaningful human-AI interactions looks at data and artificial intelligence from a creative computing angle.

**Matthias Freiberger** is a Postdoctoral Researcher at the Pioneer Centre for AI at the University of Copenhagen. His research is centered on efficient and bio-inspired artificial intelligence.

**Anders Sundnes Løvlie** is an Associate Professor and the Head of the Media, Arts, and Design group at the IT University of Copenhagen. His research is centered on experience design, humanistic HCI, museums, and artificial intelligence.

**Sebastian Risi** is a Full Professor and the co-director of the Robotics, Evolution and Art Lab at the IT University of Copenhagen. His research in artificial intelligence is centered around computational evolution to make machines more adaptive and creative.